\documentclass[twocolumn, 
prb,preprintnumbers,amsmath,amssymb,floatfix]{revtex4}

\usepackage{graphicx}
\usepackage{subfigure}
\usepackage{dcolumn}

\begin{document}

\title{Design for a Room Temperature Superconductor}
\author{W. E. Pickett}
\affiliation{Department of Physics, University of California, Davis, California, 95616}
\date{\today}
\begin{abstract}
The vision of ``room temperature superconductivity'' has appeared intermittently
but prominently in
the literature since 1964, when W. A. Little and V. L. Ginzburg began
working on the {\it problem of high temperature superconductivity} around
the same time.  Since that time the prospects for room temperature superconductivity
have varied from gloom (around 1980) to glee (the years immediately after
the discovery of HTS), to wait-and-see (the current feeling).  Recent
discoveries have clarified old issues, making it  possible to construct
the blueprint for a viable room temperature superconductor. 
\end{abstract}
\maketitle

\section{Prologue}
It is not clear when the mantra ``room temperature superconductivity'' 
first passed the lips of 
superconductivity researchers, but it has attained some prominence in the literature
since Little used it in the title (``Superconductivity at Room temperature'')
of an overview in 1965\cite{little}
of his electronic polarization mechanism proposed for polymeric
systems.  Ginzburg immediately transitioned this to two-dimensional metal-insulator
sandwich materials and generalized the description.\cite{ginzburg64}
Sad to say, these electronic-mediated superconductors are yet to be realized.
Uses of the term {\it room temperature superconductivity}
have ranged from general speculations\cite{ladik} to
glimpses of ``glitchite''\cite{anton,langer} to 
editorial coverage of public announcements\cite{jaya} 
to preplanning for applications.\cite{patel}
In the ensuing decades Vitaly Ginzburg has been the most visible and
persistent advocate that there is no theoretical justification for pessimism.
Ginzburg was elaborating more generally on 
``high temperature superconductivity'' by\cite{ginzburg67}
1967 and his optimism has never flagged.

HTS (cuprate high temperature superconductors) set the standard with
T$_c \sim$ 90 K in 1987 and rising to $\sim$130 K by the mid-1990s,
but there have been no further increases.  Since there is no viable
theory of the mechanism of HTS, there is no approach
that will allow the rational design
of new materials with higher $T_c$ in that class. 
The phonon mechanism however has an accurate microscopic theory
that is valid to reasonably strong coupling, and therefore invites the design
of materials with higher T$_c$.
The recent and very surprising breakthroughs in phonon-mediated
superconductivity can be put to use to provide a plausible recipe for the design
of a room temperature superconductor. This is the purpose of this
modest proposal, where a central feature is the emphasis on {\it control} of the
coupling, versus the standard approach of increasing the {\it brute strength}
of the coupling.

\section{Background}
Nearing the end of the 1970s there was pessimism about raising
the superconducting transition temperature (T$_c$) significantly, and
there were claims (mostly unpublished) that the maximum T$_c$ was around
30 K.  The only superconductivity known at the time (excluding the
superfluidity of $^3$He which was unlike anything seen in metals)
was symmetric ($s$-wave) pairing mediated by vibrations of the
atoms (phonons).  The maximum T$_c$ known at the time was 23 K in the A15
system.  The maximum T$_c$ had increased by only 6 K in
25 years, and it was not due to increase for several more years
(1986, when HTS was announced).

The discovery\cite{akimitsu}  in 2001 of MgB$_2$, with T$_c$=40 K, was stunning not only
because of the unimaginably high T$_c$ for phonon-mediated
superconductivity, but also because it appeared in a completely
wrong kind of material according to the accumulated knowledge.  This
aspect has been dealt with in some 
detail\cite{jan,kong,kortus,antropov,notredame} now, and the pairing
mechanism, character, and strength is well understood.
While there has been some effort to apply what has been
learned from MgB$_2$ to find other similar, or even better, superconductors,
so far MgB$_2$ remains in a class by itself.  Diamond, when it is
heavily doped with B, becomes superconducting\cite{ekimov,takano} 
with reports as high as
12 K.  B-doped diamond has features in common\cite{boeri,ucd1,blase,xiang,ma}
with MgB$_2$ except
for the two-dimensionality (2D) of MgB$_2$'s electronic structure.  It is
this 2D character we deal with here, one important aspect of which has not been
emphasized previously.  We confine our consideration to ambient 
pressure; applied pressure could well provide enhancements.
   
\section{Strong Coupling Theory}
Migdal-Eliashberg (ME) theory of superconductivity is firmly grounded 
on the material-specific level\cite{ssw} and is one of the impressive successes
of condensed matter theory.
Stripped to its basics, (phonon-coupled) T$_c$ depends on the characteristic frequency
$\Omega$ of the phonons and on the strength of coupling $\lambda$; in the 
strong coupling regime of interest here the retarded Coulomb interaction 
loses relevance.
The essence of increasing T$_c$ lies in making one or both of 
these characteristic
constants larger while avoiding structural instability.  
The focus here is on the decomposition\cite{pba} of $\lambda$
\begin{eqnarray}
 \frac{N(E_F) <I^2>}{M<\omega^2>} 	= \lambda = 
\sum_{Q}\lambda_{Q}, 
\end{eqnarray}
where the mode-$\lambda$ for momentum $Q$ is given for circular Fermi
surfaces in two dimensions by
\begin{eqnarray}
\lambda_{\vec Q} &=&
  \frac{2}{\omega_{\vec Q}N(0)}d^2_B \sum_k |{\cal M}_{k,k+Q}|^2
   \delta(\varepsilon_k)\delta(\varepsilon_{k+Q})\\ \nonumber
   &\propto& \frac{N(0)}{\omega_{Q}} d_B^2 |{\cal M}|^2
         \hat \xi(Q), \\ \nonumber
        \hat \xi(Q)&=&[N(0)]^{-2}
     \sum_k \delta(\varepsilon_k)\delta(\varepsilon_{k+Q})
       = \frac{1}{\eta_Q\sqrt{1-\eta_Q^2}}
  \label{xiQ}
\end{eqnarray}
where $\eta_Q =Q/2k_F$.
Here $N(0)$ is the Fermi level (E$_F$=0) density of states,
 $<I^2>$ is the squared electron-displaced ion matrix element
averaged over the Fermi surface, $M$ is the atomic mass, and $\omega$ is
the characteristic physical phonon frequency.  The el-ph matrix element ${\cal M}_{k,k+Q}$
involves $I_{k,k+Q}$, $M$, and $\omega$ in the standard way.  Here the band
degeneracy $d_B$ factor (number of Fermi surfaces) is treated as in MgB$_2$,
where the two $\sigma$ surfaces are equally important.

\begin{figure}[tbp]
{\resizebox{7cm}{7cm}{\includegraphics{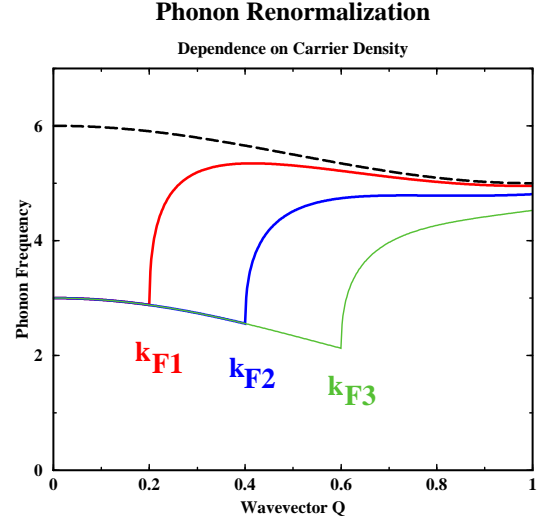}}}
\caption{
 Illustration of the renormalization of a single optic mode, shown
 as the dashed cosine-like branch, due to carriers with a 2D 
 cylindrical Fermi surface.  The downward shift (softening) for $Q<2k_F$
 is proportional to $N(E_F)<I^2>$.  There are curves for three different
 densities: lower density, $2k_{F1}$; moderate density, $2k_{F2}$;
 higher density, $2k_{F3}$.  Note in particular that the softening does
 not depend on the density ($k_F$).
      }
      \label{renorm}
      \end{figure}

In the conventional adiabatic approximation, phonon renormalization is given
by the real part of the phonon self-energy at zero frequency
\begin{eqnarray}
\Pi^{\sigma}(Q)&=&-2 \sum_{k,n,m} |{\cal M}_{k,Q}|^2
     \frac{f_{k,n} -f_{k+Q,m}}{\varepsilon_{k+Q,m}
       -\varepsilon_{k,n} } \nonumber \\
   &\approx& -2|{\cal M}|^2 d^2_B N(0) \hat{\chi}(Q),\nonumber \\
     \hat{\chi}(Q)&=&1, {\hskip 22mm} \eta_Q < 1, \nonumber \\
          &=& \bigl[1-\sqrt{1-\eta_Q^{-2}}\bigr],
	            \eta_Q > 1.
\label{chiQ}
\end{eqnarray}
Both here and above the final expression has been expressed for MgB$_2$-like
systems with cylindrical Fermi surfaces.\cite{jan,kong,2Dpickett,extreme,iim}
The behavior of this renormalization is pictured in Fig. \ref{renorm}, note
specifically the independence of the degree of softening on $k_F$.

It is the last result Eq. \ref{chiQ} that we emphasize here: 
the phonon renormalization is
constant ({\it i.e.} controlled) for $Q<2k_F$, and diminishes quickly for larger
$Q$.  For circular Fermi surfaces there is no sharp peak in $\chi(Q)$ at $2k_F$ nor
anywhere else.  In MgB$_2$, phonon coupling arises dominantly from 
$Q<2k_F$, and only the two bond-stretching branches have the very large matrix elements.

The number of phonon scattering
processes from/to the Fermi surface is quantified by the
``nesting function'' $\hat \xi(Q)$ (given here in 
normalized form whose zone sum is unity).
For the 2D circular dispersion relation, it is the simple known quantity in
Eq. \ref{xiQ}.  It has simple integrable divergences at $Q$=0 and $Q=2k_F$
which, most importantly, do not result in $Q$-dependent softening, hence avoiding
instability in spite of contributing positively to $\lambda$.  

\begin{figure}[tbp]
{\resizebox{7cm}{7cm}{\includegraphics{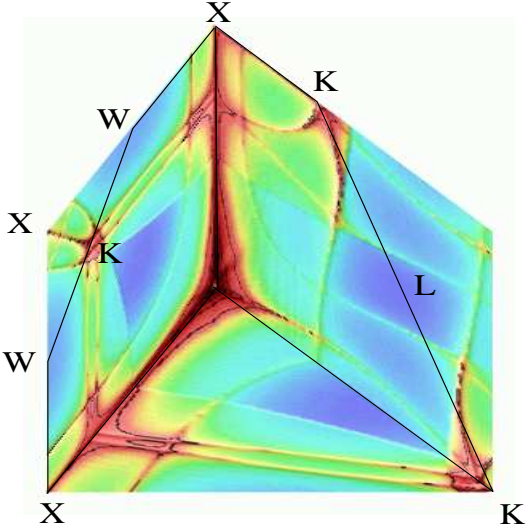}}}
\caption{ 
 Intensity plot of the nesting function $\xi(Q)$ for fcc lithium (at 36 GPa)
 in three planes in the Brillouin zone.  The dark (red) denotes high intensity,
 light denotes low intensity.  $\xi(Q)$ has very sharp structure and high
 intensities in very localized regions, in spite of the very simple (Cu-like)
 Fermi surface.  A very fine mesh of k-points (2$\times 10^6$ in the Brillouin
 zone) was used to reveal the fine structure.
 }
\label{xiQ3d}
\end{figure}

This behavior
can be contrasted with the vast uncertainty inherent in a general Fermi surface,
which has arisen forcefully in recent discoveries.  Under pressure elemental
Li metal becomes superconducting up to near 20 K as shown by three
groups,\cite{shimizu,struzhkin,schilling} in spite of being an 
$s-p$ metal with a simple Fermi surface.\cite{LiUCD,LiGross}
In Fig. \ref{xiQ3d} $\xi(Q)$ is displayed in three planes\cite{LiUCD} in the zone, where
extremely sharp structure is apparent.  The sharp structure near the K symmetry
point leads to a lattice instability (large contribution to $\hat \chi(Q)$).  
The sharp structure
occurs in spite of a very simple Cu-like Fermi surface consisting of spheres
joined by necks along $<111>$ directions.  This example demonstrates why
{\it control} of the $Q$-dependence of the coupling is essential; $\lambda$
is not so large\cite{LiUCD,LiGross} for Li at this volume ($\lambda \sim 1.5-3$) 
yet the lattice
has been pushed to instability.  This example shows that the overall value of
$\lambda$ is not the indicator of instability of the system, which occurs
when some phonon frequency is renormalized (softened) to zero.

\section{Learning from Recent Discoveries}
The crucial features to be learned  from MgB$_2$ and Li under pressure follow.\\
1.  High frequency is important; this has long been understood.  By beginning
from a very stiff unrenormalized lattice, a crystal can withstand a great deal of 
renormalization to lower frequencies that must accompany strong coupling.\\
2.  Very high mode $\lambda_Q$'s (up to 20-25 for MgB$_2$) can arise without
instability.  MgB$_2$ appears not to be near any instability, although analysis
shows\cite{notredame} that only $\sim$15-20\% stronger coupling would 
result in instability.\\
3.  The phonon softening in 2D systems due to strong coupling, and the total $\lambda$, is
independent of carrier concentration (or varies smoothly if the effective mass changes
with doping). \\ 
4.  Impressive results can be attained from strong coupling to only a fraction 
of the modes.  For MgB$_2$, T$_c$ = 40 K with only 3\% of the modes strongly
coupled.  These modes are the bond-stretch modes (2 out of 9 branches) with
$Q<2k_F$ (12\% of the zone). \\
5.  General-shaped Fermi surfaces, even the simple one of Li, can readily lead to
lattice instability from a thin surface of soft modes in Q-space.  Such instability
restricts the achievement of high T$_c$.\\
6.  Two-dimensional parabolic bands provide {\it ideal control} of the Q-dependence
of coupling strength: phonon renormalization is {\it constant} for $Q<2k_F$, not
sharply peaking in an unexpected region of the zone as can happen for a 
Fermi surface of general shape (viz. for Li
in Fig. \ref{xiQ3d}).  

\section{Putting It All Into a Design}
\subsection{What We Have}
\begin{figure}[tbp]
\rotatebox{-90}{\resizebox{8cm}{4.0cm}{\includegraphics{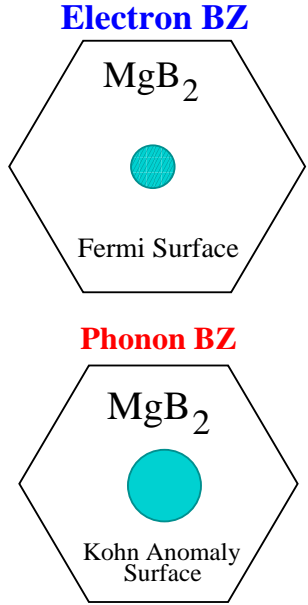}}}
\caption{
 Top: top view of the hexagonal electron Brillouin zone of MgB$_2$,
  with idealized $\sigma$-band Fermi surface circle 
   Bottom: the phonon Brillouin zone (same zone, of course), 
    illustrating the simple Kohn anomaly surface with
    radius (not diameter) $2k_F$.
       }
  \label{Hex1}
\end{figure}

The superconducting T$_c$ of MgB$_2$ is remarkable, and arises from the extreme
strong coupling of 3\% of its phonons.  The electron-phonon deformation potential
$<I^2>^{1/2}$ is very large for the bond-stretch modes, 
and MgB$_2$ gains a factor $d^2_B$=4 from having
two Fermi surfaces.  
Figure \ref{Hex1}
illustrates (1) the $\sigma$-band Fermi surfaces, idealized to identical circular
surfaces, and (2) the Kohn-anomaly region of renormalized phonons $Q<2k_F$.  
In MgB$_2$ this
``Kohn surface'' encloses only 12\% of the zone.
Previous analysis\cite{notredame} has
shown that pushing these phonons further, by either increasing the bare 
coupling strength $N(E_F)<I^2>$ or varying the lattice stiffness (bare phonon frequency)
separately, could increase T$_c$ by perhaps 20\% but then would drive the material
to a lattice instability.  It is fairly obvious that if one increases both
proportionately, then one gains by enhancing the characteristic frequency while
keeping $\lambda$ constant.  This is the {\it metallic hydrogen} scenario that
was discussed originally 40 years ago.\cite{ashcroft}

\begin{figure}[tbp]
\rotatebox{-90}{\resizebox{8cm}{4.0cm}{\includegraphics{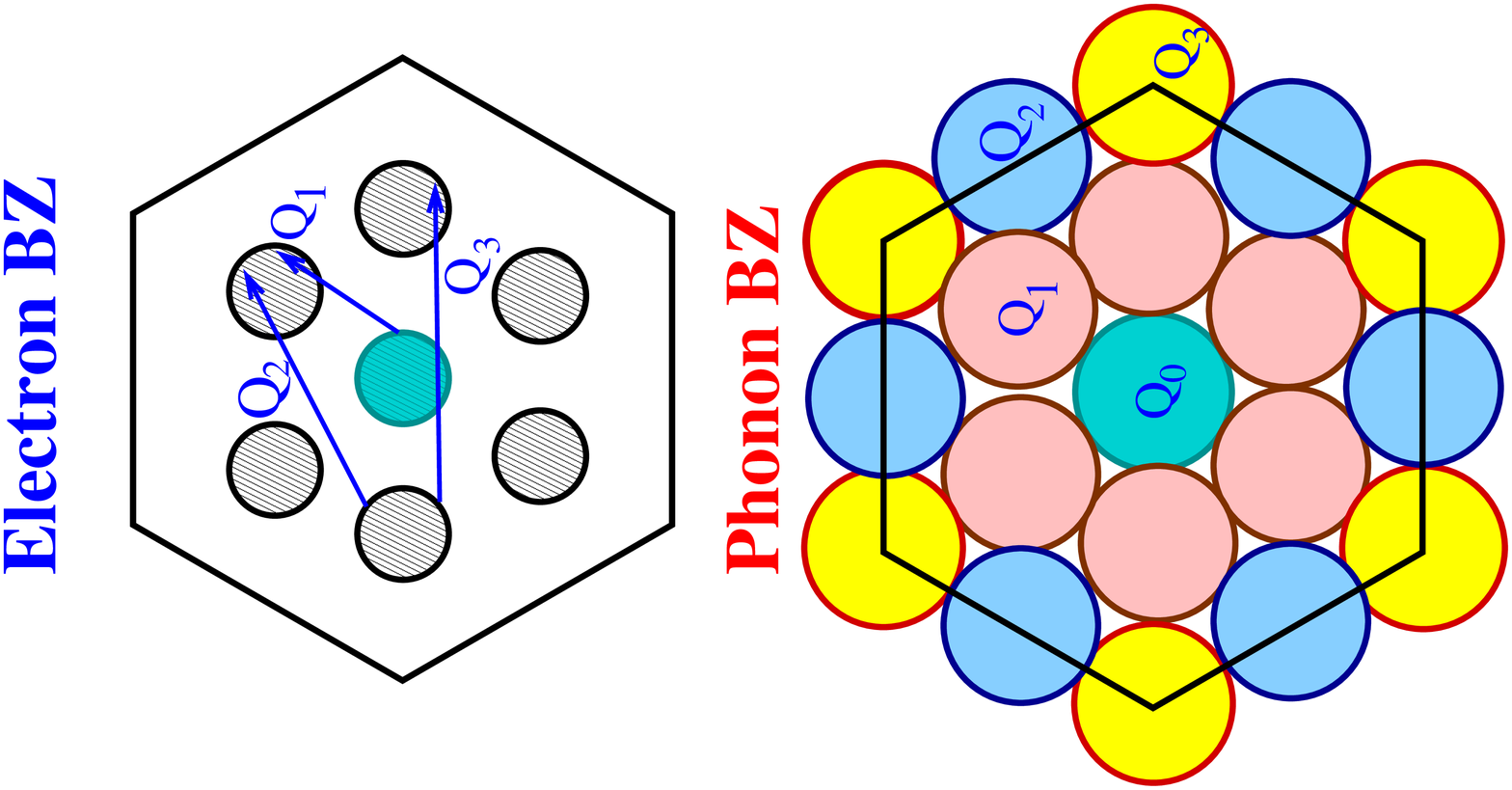}}}
\caption{
 Top: hexagonal electron Brillouin zone with central Fermi surface circle (an 
   idealization of that of MgB$_2$) and with six additional circular Fermi
   surfaces placed at the midpoint of the $\Gamma$-K line. 
 Bottom: the phonon Brillouin zone, mapping the circular regions $|Q-Q_j|<2k_F$,
   $j$=0, 1, 2, 3.  The figures are drawn for $k_F$ equal to one-eighth of the
   $\Gamma$-K line length, which results in 2D close packing of the circles
   of diameter $4k_F$.  The Kohn circles at the edge have been pictured
   extending into the neighboring zones to facilitate comparison with the
   top panel.
      }
      \label{Hex19}
      \end{figure}

\subsection{What We Need}
There is however the equally obvious approach: introduce more phonons into the
coupling.  From this viewpoint, MgB$_2$ with its 3\% of useful phonons seems
pathetic, yet as noted in Sec. III, increasing the number of strongly 
coupled phonons be increasing the doping does not enhance $\lambda$ because
the individual mode $\lambda_Q$'s decrease accordingly.  And further
increasing the matrix elements, which increases $\lambda$, also readily
leads to instability.

The key to increasing $\lambda$ is not in doing more (more carriers, hence more
phonons), but in doing it {\it again} and {\it again}, with other Fermi
surfaces and other phonons.
Here the {\it control}, not over the strength of the coupling  but over
the {\it specific momenta} [that is, $\lambda_Q$, or more specifically $\chi(Q)$] 
becomes essential.  The requirement is to renormalize frequencies in other parts
of the zone {\it without further softening} the ones that are already very
strongly renormalized, since that will not lead to lattice instability.  
Two-dimensional
circularly-symmetric dispersion relations provide that control, as demonstrated
by the equations above.  Adding an additional circular Fermi surface centered at
$Q_1$ gives another region of renormalized phonons, and contributions $\lambda_Q$
to $\lambda$, of radius $2k_F$ centered at $Q_1$. 

Let us skip directly to the optimal case, illustrated in Fig. \ref{Hex19}.  By
adding to the MgB$_2$ zone-centered surface another spherical Fermi surface 
half-way along the $\Gamma$-K line,
with radius 1/8 of the length of the $\Gamma$-K line, one obtains three new
spanning vectors $Q_1, Q_2, Q_3$ and their symmetry partners that produces
an array of close-packed Kohn surfaces of radius $2k_F$ within which coupling
is strong.  Given the close-packing
fraction in 2D, $\pi/2\sqrt{3}$=0.907, this arrangement manages to
utilize 90\% of the zone for strongly coupled phonons, an increase in fraction
of zone used, compared to MgB$_2$ (12\%), by a factor of 7.5. 

This extension from MgB$_2$ is not yet optimum, because MgB$_2$ uses only 2/9
of its branches.  Optimally, every branch would be drafted into 
service in strong coupling, giving another factor of 4.5, or a total 
enhancement of $\sim$30-35.
The strongly coupled modes in MgB$_2$ have mode-$\lambda_Q$'s of mean value 20-25
(calculations so far have not been precise enough to pin this down).  Let us
not be pessimistic, and therefore use $\bar \lambda_Q$=25; this value is consistent
with the value of $\lambda$ from the strongly coupled modes divided by 3\%,
i.e. 0.7/0.03$\sim$23.  Then  with 
90\% participation of the phonons $\lambda$=25$\times 0.90$=22.5.  
Using the Allen-Dynes equation\cite{alldyn}
to account properly for the strong-coupling limit that is being approached,
and using the MgB$_2$ frequency of 60 meV, one obtains T$_c$=430 K.  The strong
coupling limit\cite{carbotte} of the ratio $2\Delta/k_B T_c$ is 13, 
so we can estimate the gap of
such a superconductor to be $2\Delta \sim 12 k_B T_c \sim 0.4$ eV.
This will be an interesting superconductor indeed.

\subsection{Where To Look}
Whereas the architectural design of a room temperature superconductor provided
here is straightforward, the structural engineering necessary to implement 
this vision will 
require creativity and knowledge of materials.  Where does one look?

\subsubsection{More Fermi Surfaces}
Figure \ref{Hex19} gives an idea of a set of Fermi surfaces that could be
promising.  This pattern is very much like that of Na$_x$CoO$_2$, although
its T$_c$=4.5 K, and even that not necessarily from phonons, is not a good
example of strong coupling enhanced by Fermi surface arrangement.  (The
zone-centered Fermi surface in Na$_x$CoO$_2$ is much larger than the others.)  The
encouraging feature to be emphasized here is that this example 
shows that the desirable arrangement
of Figure \ref{Hex19} is nothing unusual.  Not only are the Fermi
surfaces very near the midpoint
of the $\Gamma$-K line, but they are nearly circular although not enforced by
symmetry.

A more provocative example is that of the layered, electron doped system
Li$_x$MNCl, which has up to T$_c$=15 K for M=Zr, and T$_c$=26 K for M=Hf.
For $0 < x < 0.4$ which probably includes the accessible concentrations, this
system has very nearly circular K-centered Fermi surfaces whose nesting possibilities
have been noted.\cite{weht,heid}  There are surfaces at two inequivalent
points K, resulting in Kohn regions centered at the zone center (degeneracy two)
and [because K$\rightarrow$K scattering involves momentum transfer K in the 
hexagonal zone]
at each of the pair of points K (also degeneracy two).  The figure of Kohn surfaces
would look like Fig. \ref{Hex19} with circles centered at the $Q_1$ and $Q_2$
points missing.  Indeed, Heid and Bohnen calculate\cite{heid} phonon softening 
around K and the zone
center, but alas they find that the calculated coupling strength
comes well short of accounting for T$_c$=15 K in Li$_x$ZrNCl (not to mention the
question of T$_c$=26 K in Li$_x$HfNCl).  The superconductivity in this system 
remains unexplained, but the placement of its 2D circular Fermi surfaces
illustrates the first step in pulling more phonons into the coupling.

\subsubsection{More Phonon Branches}
While the electronic system we seek is 2D in character, we also seek (strong)
coupling to phonon of arbitrary momentum and polarization.  MgB$_2$ is a
bad example here, since the strong bonding lies solely within the layer;
the $\hat z$-axis modes are hardly coupled.  A lattice with strong
bonds with substantial component perpendicular to the layers have a better
likelihood of coupling to bond stretch modes.  From this point of view the
Li$_x$ZrNCl system seems to be a step in the right direction, as it has
Zr-N bonds both within the layer and perpendicular to it.  
Heid and Bohnen\cite{heid}
found moderate coupling both to two stiff modes ($\sim$65 meV, primarily N) 
and to two softer modes
($\sim$25 meV).  These modes are however polarized in-plane; the perpendicular
Zr-N modes do not couple to the 2D electronic system.  Creativity will be
required to find how to involve a larger fraction of modes in the coupling.

\section{Remaining Issues}
Since there may be people who are truly interested in this topic of room 
temperature superconductivity, the basic premises and concepts should be made
as clear and precise as possible (``truth in advertising'').  The following
issues should be mentioned.\\
1.  The use of the term {\it optimal} as used above was disingenuous and
incorrect, and that discussion as it stands
is unduly pessimistic.  Close-packing of $4k_F$-diameter Kohn
circular surfaces does indeed use the area of the zone efficiently (although
mathematicians or engineers would fill the holes with smaller circles, and then
again and again ad infinitum, thereby using 100\% of the area).  However,
maximizing the area is {\it not} the objective.  The objective is to maximize
T$_c$ or to simplify slightly for now, to maximize $\lambda$.  Phase space in 2D
leads to the non-intuitive result that integration of $\lambda_Q$ over a Kohn
region of diameter $4k_F$ is independent of $k_F$, as long as the ME theory we are
applying holds.  The objective then is to maximize the {\it number} of Kohn
regions.  Since there are likely to be different values of $N(0)<I^2>$ and $\omega$
for the different regions, it will be something like the quantity
\begin{eqnarray}
 \sum_j^{{\cal N}_K} \frac{ [N(0)<I^2>]_j}{\omega_j}
 \label{optimum}
\end{eqnarray}
that needs maximizing, where ${\cal N}_K$ is the (variable)
number of Kohn regions.\\
2. Care must be taken to keep the Kohn regions from overlapping if the combined
renormalization of phonons will be so strong as to drive instability.  Also, the band
degeneracy factor $d_B$ seems to be extremely helpful, but must be watched. (In
Eq. \ref{optimum} there would be some factor related to $d_B$ of identical terms.)
For example, in Fig. \ref{Hex19} each of the Fermi surfaces contributes (via
intrasurface scattering) to
the coupling strength and accompanying phonon softening inside the Kohn region at
$Q_0 = 0$, a factor of 12 in this case.  A good strategy would be to have 
matrix elements be larger for $Q_1, Q_2, Q_3$, which have smaller degeneracy
factors, than for $Q_0$ with its large multiplier, thereby crafting strong
coupling while avoiding the $Q=Q_0$ instability.\\
3. The validity of the theory bears reconsideration.  It is safe to say that if
$\lambda_Q (\omega_Q/E_F) \ll 1$ for every phonon $Q$, then ME (second order
perturbation) theory is safe.  If this inequality is not satisfied for a small 
fraction of the coupling strength, corrections are probably minor.  However, for the
typical strongly-coupled phonon in MgB$_2$,
\begin{eqnarray}
\lambda_Q \frac{\omega_Q}{E_F} \sim 25 \frac{60 meV}{400 meV} \sim 4.
\end{eqnarray}
The condition of validity is violated badly for {\it every important phonon}, making ME
theory in MgB$_2$ unjustified as pointed out 
earlier.\cite{lilia1,brazil,lilia2}  Nevertheless,
its predictions seem to be reasonable for MgB$_2$, so it is reasonable for us
to extrapolate the theory to include more phonons with similar strength of
coupling.\\
4. It was noted above that for a frequency $\omega\sim$ 60 meV as in MgB$_2$, 
$\lambda \sim$ 20 is required to reach the vicinity of room temperature.  While
there have been many papers addressing the very strong regime and resulting 
polarons and bipolarons, we are not aware of any that address seriously such
coupling in a degenerate electron system.  There is a general expectation that
the electron system becomes unstable, but unstable to what is unclear; a 
degenerate gas of polarons (of the order of one per atom) does not seem like a
clear concept.  (If the instability occurs only below T$_c$, it may not control
or limit the pairing at high temperature.)   The description of such a 
system is still unknown; while MgB$_2$
has a finite concentration (3\%) of {\it extremely} strongly coupled phonons, the net value
of $\lambda$ is less than unity.  Following the materials design proposed here
will lead to study of a new materials regime as well as higher T$_c$.

\section{Prospects for Success}
A blueprint for the design of a room temperature superconductor has been provided
here.  The essence is this: very strong Q-dependent coupling in a {\it controlled}
fashion is possible, and one follows the path toward getting as much as 
one can out of both the electronic
and the phononic systems.  Two-dimension electronic structures with circular
Fermi surfaces give an unsurpassed level
of control of the Q-dependence, which if uncontrolled can and does lead to
structural instabilities even at moderate total coupling strength.  
Stiff lattices are important, as is getting as many
of the branches as possible involved in coupling.

On the one hand MgB$_2$, impressive as it is, seems to be doing a poor job
in most respects of making use of the available phase space.  MgB$_2$ uses
only 12\% of all the possible phonon momenta, and only 2/9 of its phonon
branches, netting only 3\% of phonons involved in coupling.  MgB$_2$ excels at
producing extremely large electron-displaced ion matrix elements, and its
2D phase space keeps the extremely large coupling (to those 3\%)
firmly under control.  All things considered, it seems reasonable to expect
that materials exist, or can be made, that will improve on MgB$_2$'s current
record.  

There is no promise here, nor even expectation, that such improvement will
be easy.  Although one can employ tight-binding models to suggest 
crystal structures and interatomic interactions that will place band
extrema in the desired positions in the zone, synthesizing the corresponding
material is more uncertain.  We work with discrete nuclear charges, so
this is a Diophantine problem rather than a continuous one, and bonding
properties can change rapidly from atom to atom.  In addition, incremental
improvements lead to even more incremental payoffs.
The behavior\cite{alldyn} of T$_c$($\lambda$) $\propto \sqrt{\lambda}$ 
in the strong-coupling regime,
while monotonically increasing, provides a law of diminishing returns:
doubling T$_c$ requires quadrupling $\lambda$, a sobering prospect.
Finally, it should re-emphasized that no realistic theory of the
degenerate electronic system in the presence of phonon coupling in the
regime $\lambda > 5$ (say) exists, but this is another, perhaps better,
physical reason to push to stronger coupling systems.
And think of it --- wouldn't it be really great to carry around a spool of
superconducting D$_2$E$_3$Z wire in your pocket?
\section{Acknowledgments}
This work was stimulated in part by preparation for, and attendance of, the
workshop on {\it The Possibility of Room Temperature Superconductivity},
held June 2005 at the University of Notre Dame.  I have benefited from
discussion and collaboration with numerous workers in the field. 
Our recent work in this area has been supported by
National Science Foundation Grant
DMR-0421810.  Support from the Alexander von Humboldt Foundation during the latter
part of this work is 
gratefully acknowledged.


\begin{thebibliography}{10}
\bibitem{little}W. A. Little, Scientific American {\bf 212}, 21 (1965).
\bibitem{ginzburg64}V. L. Ginzburg, Phys. Lett. {\bf 13}, 101 (1964);
Sov. Phys. JETP {\bf 19}, 269 (1964). {\it On the Problem of High Temperature
 Superconductivity}
\bibitem{ladik}J. Ladik and A. Bierman,  Phys. Lett. A {\bf 29}, 636 (1969).  
 {\it On the Possibility of Room-Temperature Superconductivity in 
      Double Stranded DNA}
\bibitem{anton}K. Antonowicz, Nature {\bf 247}, 358 (1974).
  {\it Possible Superconductivity at Room Temperature}
\bibitem{langer}J. Langer, Solid State Commun. {\bf 26}, 839 (1978).
 {\it Unusual Properties of Aniline black -- Does Superconductivity
  Exist at Room-Temperature?}
\bibitem{jaya}K. S. Jayaraman, Nature {\bf 327}, 357 (1987).
 {\it Superconductivity at Room-Temperature}
\bibitem{patel}C. K. N. Patel and R. C. Dynes, Proc. Natl. Acad. Sci. {\bf 85},
 4945 (1988). 
 {\it Towards Room-Temperature Superconductivity}
\bibitem{ginzburg67}V. L. Ginzburg and D. A. Kirzhnitz, Doklady Akad. Nauk. SSSR {\bf 176},
   553 (1967). {\it On High-Temperature and Surface Superconductivity}

\bibitem{akimitsu}J. Nagamitsu,
  N. Nakagawa, T. Muranaka, Y. Zenitani, and
  J. Akimitsu,
  Nature {\bf 410}, 63 (2001).

\bibitem{jan}J.M.\ An and W.E.\ Pickett, Phys.\ Rev.\ Lett.\ {\bf 86}, 4366
   (2001)
\bibitem{kong}Y.\ Kong, O. V. Dolgov, O. Jepsen, and O. K. Andersen,
      Phys. Rev. B {\bf 64}, 020501 (2001).
\bibitem{kortus}J.\ Kortus, I. I. Mazin, K. D. Belashchenko, V. P.
        Antropov, and L. L. Boyer, Phys.\ Rev.\ Lett.\ {\bf 86},
	   4656 (2001).
\bibitem{antropov}I. I. Mazin and V. P. Antropov, Physica C
	        {\bf 385}, 49 (2003).

\bibitem{notredame}W. E. Pickett, submitted to Proc. of Notre Dame
		  conference.

\bibitem{ekimov}E. A. Ekimov,
  V. A. Sidorov, E. D. Bauer, N. N. Mel'nik,
  N. J. Curro, J. D. Thompson, and S. M. Stishov,
  Nature {\bf 428}, 542 (2004).
\bibitem{takano} Y. Takano, M. Nagao, K. Kobayashi, H. Umezawa, I. Sakaguchi,
  M. Tachiki, T. Hatano, and H. Kawarada,
    Appl. Phys. Lett. {\bf 85}, 2581 (2004).
\bibitem{boeri} L. Boeri, J. Kortus, and O. K. Anderson,
  Phys. Rev. Lett. {\bf 93}, 237002 (2004).
\bibitem{ucd1} K.-W. Lee and W. E. Pickett,
  Phys. Rev. Lett. {\bf 93}, 237003 (2004).
\bibitem{blase} X. Blase, Ch. Adessi, and D. Conn\'etable,
  Phys. Rev. Lett. {\bf 93}, 237004 (2004).
\bibitem{xiang} H. J. Xiang, Z. Li, J. Yang, J. G. Hou, and Q. Zhu,
   Phys. Rev. B {\bf 70}, 212504 (2004).
\bibitem{ma}Y. Ma, J. S. Tse, T. Cui, D. D. Klug, L. Zhang, Y. Xie, Y. Niu,
	       and G. Zou, Phys. Rev. B {\bf 72}, 014306 (2005).
		

\bibitem{ssw}D. J. Scalapino, J. R. Schrieffer, and J. W. Wilkins,
  Phys. Rev. {\bf 148}, 263 (1966).

\bibitem{pba}P. B. Allen, Phys. Rev. B {\bf 6}, 2577 (1972);
 P. B. Allen and M. L. Cohen, Phys. Rev. Lett. {\bf 29}, 1593 (1972).
 A numerical correction is given in Eq. (4.27) of P. B. Allen, in
  {\it Dynamical Properties of Solids}, Ch. 2, edited by G. K. Horton and
  A. A. Maradudin (North-Holland, Amsterdam, 1980).


\bibitem{2Dpickett}W. E. Pickett, J. M. An, H. Rosner, and S. Y. Savrasov,
  Physica C {\bf 387}, 117 (2003).
\bibitem{extreme}J. M. An, S. Y. Savrasov, H. Rosner, and W. E. Pickett,
  Phys. Rev. B {\bf 66}, 220502 (2002).
\bibitem{iim}I. I. Mazin, private communication.

\bibitem{shimizu}K. Shimizu, H. Kimura, D. Takao, and K. Amaya,
   Nature {\bf 419}, 597 (2002).
\bibitem{struzhkin}V. V. Struzhkin, M. I. Eremets, W. Gan, H.-K. Mao,
   and R. J. Hemley, Science {\bf 298},
   1213 (2002).
\bibitem{schilling}S. Deemyad and J. S. Schilling, Phys. Rev. Lett.
   {\bf 91}, 167001 (2003).

\bibitem{LiUCD}D. Kasinathan, J. Kune\v{s}, A. Lazicki,
   H. Rosner,
   C. S. Yoo, R. T. Scalettar, and W. E. Pickett, Phys. Rev. Lett.
   {\bf 96}, 047004 (2006).
\bibitem{LiGross}G. Profeta, C. Franchini, N. N. Lathiotakis,
  A. Floris, A. Sanna, M. A. L. Marques, M. Lueders, S. Massidda,
   E. K. U. Gross, and A. Continenza, Phys. Rev. Lett.
   {\bf 96}, 047003 (2006).

\bibitem{ashcroft}N. W. Ashcroft, Phys. Rev. Lett. {\bf 21}, 1748 (1968).

\bibitem{alldyn}P. B. Allen and R. C. Dynes, Phys. Rev. B
{\bf 12}, 905 (1975).
\bibitem{carbotte}J. P. Carbotte, Rev. Mod. Phys. {\bf 62}, 1027 (1990).

\bibitem{LiBC}H. Rosner, A. Kitaigorodsky, and W. E. Pickett,
  Phys. Rev. Lett. {\bf 88}, 127001 (2002).




\bibitem{weht}R. Weht, A. Filippetti, and W. E. Pickett, Europhys. Lett.
  {\bf 48}, 320 (1999).
\bibitem{heid}R. Heid and K.-P. Bohnen, Phys. Rev. B {\bf 72}, 134527 (2005).

\bibitem{brazil}W. E. Pickett, Braz. J. Phys. {\bf 33}, 695 (2003).
\bibitem{lilia1}L. Boeri, G.B. Bachelet, E. Cappelluti and L. Pietronero,
  Phys. Rev. B {\bf 65}, 214501 (2002).
\bibitem{lilia2} L. Boeri, E. Cappelluti and L. Pietronero,
  Phys. Rev. B {\bf 71}, 012501 (2005).

\end{thebibliography}
\end{document}